\documentclass[twocolumn,showpacs,preprintnumbers]{revtex4}%
\usepackage{amsfonts}
\usepackage{amsmath}
\usepackage[dvips]{graphicx}
\usepackage{dcolumn}
\usepackage{bm}%
\usepackage{amssymb}
\begin{document}
\preprint{ }
\title{Search for a dark matter particle in high energy cosmic rays}
\author{Yukio Tomozawa}
\affiliation{Michigan Center for Theoretical Physics, Randall Laboratory of Physics,
University of Michigan, Ann Arbor, MI. 48109-1040}
\date{\today }

\begin{abstract}
Existing data hints that high energy cosmic ray experiments may offer the most
promissing shot at finding a dark matter particle. A search in the PeV mass
range is suggested, where the discovery of such a particle might help explain
the GZK cutoff violation data.

\end{abstract}

\pacs{04.70.-s, 95.85.Pw, 95.85.Ry, 98.54.Cm}
\maketitle

\section{Introduction}

An extensive search for a dark matter particle (DMP) is under way throughout
the world\cite{dmsearch} by underground detectors using cryogenic or
electronic methods. However, there is no observational hint whatsoever as to
the mass and interactions etc. of the particle searched for. A slight hope is
that the forthcoming LHC experiment might give some hint as to the nature of
the particle. Such a possibility might be wishful thinking, in view of the
absence of any hint from Tevatron experiments in the few TeV\ energy range.
The author will try to construct a scenario for a DMP search, as much as
possible based on the observational data.

\section{The AGASA data on the GZK cutoff violation}

High energy cosmic rays traversing intergalactic space suffer the GZK
cutoff\cite{gzk} above 100 EeV due to interactions with cosmic background
radiation, if the primary cosmic ray particles are protons or nuclei. The
Pierre Auger Project\cite{auger}, HiRes\cite{hires} and Yakutsk\cite{yakutsk}
found the GZK cutoff, while Akeno-AGASA\cite{agasa} observed the events above
the cutoff (11 events in the past 10 years)\cite{puzzle}. Since the number of
events that violate the GZK cutoff has been steadily increasing in the past 10
years, the discrepancies among the results for different detectors must be
explained by experimentalists. Since the result of the Akeno-AGASA experiment
is smooth near the cutoff energy, we have to accept their result and wait for
a future explanation of the differences among the detectors. The author will
assume that the Akeno-AGASA result is correct and consider its implication,
until otherwise claimed.

A possible explanation for the AGASA data on GZK cutoff violation would be a
shower caused by a DMP. A DMP is not constrained by the GZK cutoff, since it
iteracts weakly with cosmic background radiation. Then the question is how
such a particle can be accelerated to an energy as high as 100 EeV. This is
intimately related to the question of what is the source of high energy cosmic
rays. Recent measurements by the Pierre Auger Project on the correlation
between the direction of high energy cosmic rays and the location of AGN
(Active Galactic Nuclei)\cite{auger} will shed some light on this question.
The crucial question is then how any particle can be accelerated to as high as
100 EeV in connection with AGN. This is precisely the question that the author
confronted since 1985. A model that the author presented in 1985 predicted the
data from the Pierre Auger Project and at the same time solved the problem of
accelerating a dark matter particle, which is a neutral particle. We start
with a review of that model.

\section{High energy cosmic rays from AGN and existence of a new particle
around the PeV mass range}

In a series of articles\cite{cr1}-\cite{cr9}, the author has presented a model
for the emission of high energy particles from AGN. The following is a summary
of the model.

1) Quantum effects on gravity yield repulsive forces at short
distances\cite{cr1},\cite{cr3}.

2) The collapse of black holes results in explosive bounce back motion with
the emission of high energy particles.

3) Consideration of the Penrose diagram eliminates the horizon problem for
black holes\cite{cr4}. Black holes are not black anymore.

4) The knee energy for high energy cosmic rays can be understood as a split
between a radiation-dominated region and a matter dominated region, not unlike
that in the expansion of the universe. (See page 10 of the lecture
notes\cite{cr1}-\cite{cr3}.)

5) Neutrinos and gamma rays as well as cosmic rays should have the same
spectral index for each AGN. They should show a knee energy phenomenon, a
break in the energy spectral index, similar to that for the cosmic ray energy spectrum.

6) The recent announcement by Hawking rescinding an earlier claim about the
information paradox\cite{hawking} is consistent with this model.

Further discussion of the knee energy in the model yields the existence of a
new mass scale in the knee energy range, in order to have the knee energy
phenomenon in cosmic ray spectrum\cite{crnew}. The following are additional
features of the model.

7) If the proposed new particle with mass in the knee energy range (0.1
PeV$\sim$2 PeV) is stable and weakly interacting with ordinary particles, then
it becomes a candidate for a DMP. It does not necessarily have to be a
supersymmetric particle. That is an open question. However, if it is
supersymmetric, then it is easy to make a model for a weakly interacting
DMP\cite{pevss}. The only requirement is that such particles must be present
in AGN or black holes so that the the knee energy is observed when cosmic rays
are emitted from AGN. A suggested name for the particle is sion
(xion)\cite{crnew}, using the Chinese/Japanese word for knee, si (xi).

8) If the particle is weakly interacting, then it does not obey the GZK
cutoff, since its interaction with photons in cosmic backgroud radiation is
weak, as was pointed out earlier. This is a possibe resolution of the GZK puzzle.

In summary, this model predicts the Pierre Auger Project data. Moreover it
suggests the existence of a new particle in the PeV mass range, in order to
explain the knee energy phenomenon of cosmic ray spectrum.

\section{ Search for a new particle by high energy cosmic ray detectors.}

We assume that the incident particles above the GZK cutoff observed by the
Akeno-AGASA detector are weakly interacting particles at the PeV mass scale,
which are required to exist in order to explain the phenomenon of the cosmic
ray knee energy in the model. In a model where the acceleration takes place by
gravity such as that proposed by the author, there is no difficulty in
accelerating a weakly interacting and neutral DMP. One has to explain a
mechanism whereby the Akeno-AGASA detector is sensitive to such weakly
interacting particles and all other detectors are not, as was pointed out
earlier. This is quite conceivable, since the spacing of the detectors in the
Akeno-AGASA apparatus is small (1 km between detectors) compared with that of
the other detectors (1.5 km between detectors for the Pierre Auger Project).
Besides, a direct measurement of showers may give higher sensitivity for
weakly interacting particles: Since weakly interacting particles in high
energy cosmic rays tend to make showers at lower altitude in the atmosphere
due to the smaller cross sections, the Akeno-AGASA detector is expected to
observe a higher percentage of weakly interacting particles. Leaving this task
of quantitative estimate to the experimentalists, the author suggests the
following method for detecting such a DMP.

A weakly interacting DMP has interactions increasing with energy, similar to
the ordinary weak interactions of the standard model. The strength reaches a
maximum at an energy comparable to the mass scale, i.e., at PeV center of mass
energy. This corresponds to a lab energy of 100 EeV. Thus, a weakly
interacting DMP can make a shower, maybe at a lower level of the atmosphere.
If it is a sion, it has to be produced abundantly in AGN, to the extent that
it produces the knee energy phenomenon in cosmic ray energy spectrum. If it is
a neutral component of a sion, a collision with an atmospheric nucleus has to
produce the same particle carrying a significant fraction of the initial
momentum along with a shower, but it does not contribute to the production of
a shower. Such a particle can produce a secondary shower in the neighborhood
of the center of the primary shower after passage through the atmosphere or it
may completely disappear from the sight. It can produce an artificial shower
from shielding material in front of, say the muon detectors. This
consideration naturally yields a possible scenario for the detection of a DMP.
See Fig. 1 for a schematic layout of experimental setup.

1) Choose a cosmic ray shower detector equipped with underground muon
detectors. With muon detectors at an appropriate depth, the thickness of the
earth may play the role of shield material. If not, one has to provide some
thickness of shielding in front of the secondary muon detectors. An artificial
shower in the shield-muon detector system coincident with the primary shower
and near its center constitutes evidence for a DMP. Some of the existing
cosmic ray shower detectors, Auger\cite{auger}, hires\cite{hires},
yakutsk\cite{yakutsk}, AGASA\cite{agasa}, tibet\cite{tibet}, dice\cite{dice},
cacti\cite{cacti}, hegra\cite{hegra}, kascade\cite{kascade}, may be utilized
for such a purpose with small modifications. If the incident particles are
protons or nuclei, there is no secondary shower observed, unless much higher
energy than 100 EeV are attained, where pair production of PeV DMP is expected.

2) Collect the shower energy data. The threshold value for the energy of the
secondary shower provides the lowest mass value for a DMP. Find the energy
dependence of the fraction of the DMP component in high energy cosmic rays.
Extend the measurement above the GZK cutoff energy, 100 EeV.

3) Find whether measurements with the Akeno-AGASA equipment can relate the
data above\ and below the GZK cutoff. This may establish the weakly
interacting nature of the particle that causes the secondary shower and at the
same time resolve the puzzle of the GZK cutoff violation\cite{puzzle}.

4) Study the difference in nature between showers caused by nuclear particles
and those caused by a DMP. That will be useful for confirmation of the
detection of a DMP.

5) Large muon detectors in collider experiments may be combined with nearby
cosmic ray shower detectors in the search for a DMP. Alternatively, they can
be utilized as independent detectors for a DMP shower, if the nature of a DMP
shower is clarified.

The advantage of this method for a DMP search is that all the arguments are
based on observational data, compared with existing DMP searches. Due to the
nature of weakly interacting particles, the interaction cross sections
increase with energy and therefore high energy cosmic rays are a natural place
to look for DMP. Recent data of the Pierre Auger Project along with the
prediction of the author's model of the author suggests that new particles,
sions, must be produced abundantly in the AGN.

\begin{acknowledgments}
\bigskip The author would like to thank Lawrence W. Jones and Jean Krisch for
useful discussion and David N. Williams for reading the manuscript.
\end{acknowledgments}

\bigskip

Figure caption

\bigskip

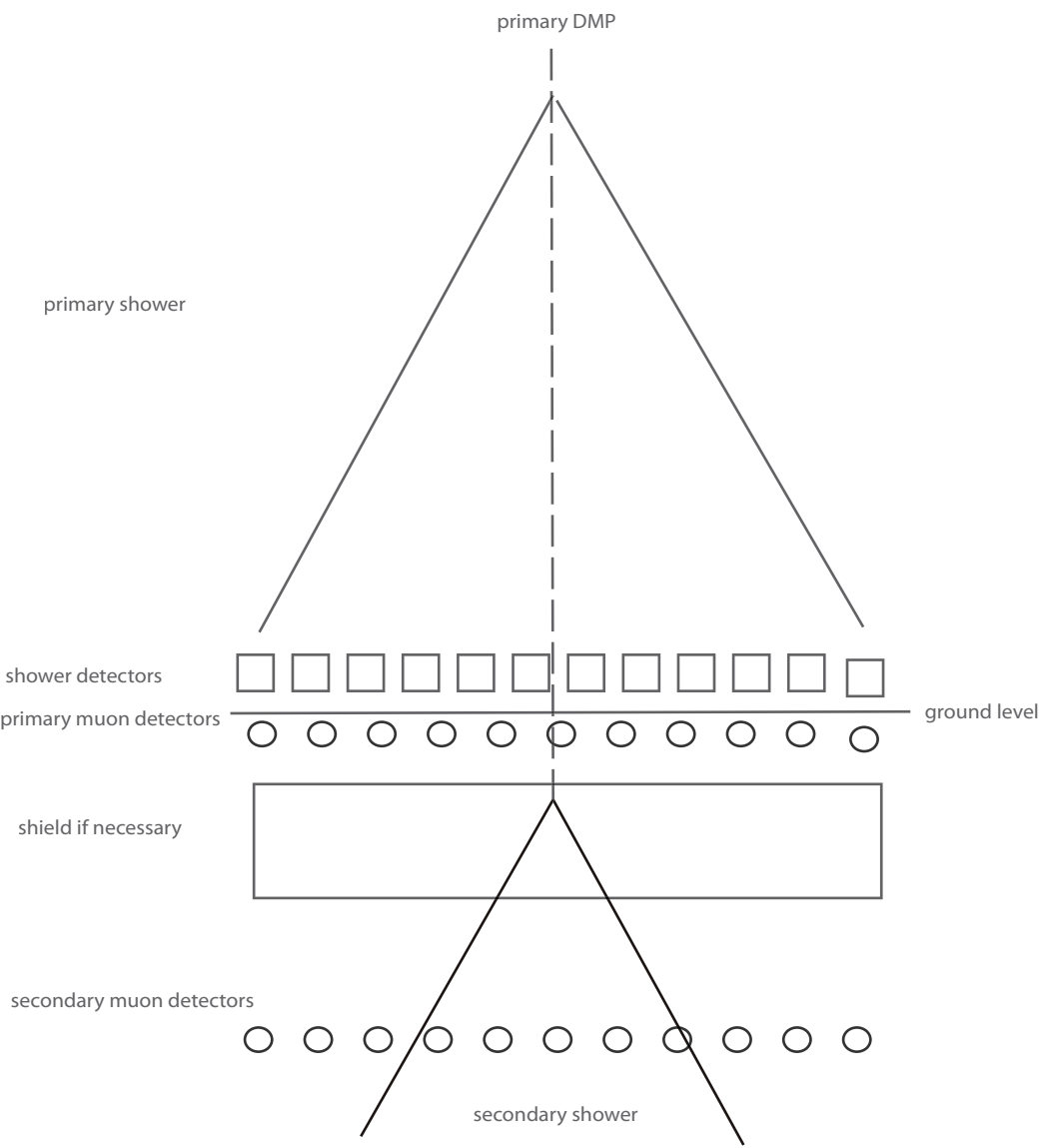
Fig. 1 Schematic figure for dark matter particle search in cosmic ray experiment.

\end{document}